\documentclass[aps,preprintnumbers,twocolumn]{revtex4}

\usepackage{graphicx}
\usepackage{bm}
\renewcommand{\Re}{{\cal R}e}
\renewcommand{\Im}{{\cal I}m}

\newcommand{\phicp}{\phi_{\text{CP}}}
\newcommand{\amu}{a_{\mu}}
\newcommand{\dmu}{d_{\mu}}
\newcommand{\de}{d_e}
\newcommand{\amusm}{a_{\mu}^{\text{SM}}}
\newcommand{\amunp}{a_{\mu}^{\text{NP}}}
\newcommand{\dmusm}{d_{\mu}^{\text{SM}}}
\newcommand{\dmunp}{d_{\mu}^{\text{NP}}}
\newcommand{\amuexp}{a_{\mu}^{\text{exp}}}

\newcommand{\msel}{m_{\tilde{e}}}
\newcommand{\msmu}{m_{\tilde{\mu}}}
\newcommand{\tb}{\tan\beta}
\newcommand{\gsim}{ \mathop{}_{\textstyle \sim}^{\textstyle >} }

\newcommand{\ecm}{e~\text{cm}}

\newcommand{\gev}{\text{GeV}}
\newcommand{\tev}{\text{TeV}}

\newcommand{\eg}{{\em e.g.}}

\newcommand{\eqref}[1]{Eq.~(\ref{#1})}

\newcommand{\bold}[1]{{\text{\normalsize\boldmath $#1$}}}

\begin{document}

\preprint{MIT--CTP--3197, UCI--TR--2001--30,
          CERN--TH/2001--270,
          hep-ph/0110157,
          Snowmass P3-07}

\title{Muon Dipole Moment Experiments: Interpretation and Prospects}



\author{Jonathan L.~Feng}
\email[]{jlf@mit.edu}
\affiliation{Center for Theoretical Physics,
             Massachusetts Institute of Technology,
             Cambridge, MA 02139, USA}
\affiliation{Department of Physics and Astronomy, 
             University of California, Irvine, CA 92697, USA}
\author{Konstantin T.~Matchev}
\email[]{Konstantin.Matchev@cern.ch}
\affiliation{Theory Division, CERN,
             CH--1211, Geneva 23, Switzerland}
\author{Yael Shadmi}
\email[]{yshadmi@physics.technion.ac.il}
\affiliation{Department of Physics,
             Technion, Technion City,
             32000 Haifa, Israel}

\date{October 11, 2001}

\begin{abstract}
We examine the prospects for discovering new physics through muon
dipole moments. The current deviation in $g_{\mu}-2$ may be due
entirely to the muon's {\em electric} dipole moment. 
We note that the precession frequency in the proposed BNL muon EDM
experiment is also subject to a similar ambiguity, but this can be
resolved by up-down asymmetry measurements.  We
then review the theoretical expectations for the muon's electric
dipole moment in supersymmetric models.
\end{abstract}

\maketitle

\section{Introduction}
\label{intro}

The Standard Model of particle physics provides an extremely
successful description of all known particles and their interactions,
but fails to address many deeper questions concerning their physical
origin.  Among the least understood phenomena is CP violation.  At
present, the only observed source of CP violation in the Standard
Model is the phase of the CKM matrix.  Its fundamental origins are
unknown.  Further, while CP violation is an essential ingredient of
almost all attempts to explain the matter-antimatter asymmetry of the
universe~\cite{Sakharov:1967dj} (alternative explanations are subject
to stringent bounds: see, \eg, Ref.~\cite{Cohen:1998ac}), the amount
of CP violation present in the CKM matrix is insufficient to explain
the observed asymmetry~\cite{asymmetry}.  Searches for CP violation
beyond the CKM matrix are necessary to shed light on this puzzle and
are also probes of physics beyond the Standard Model.

Electric dipole moments (EDMs) violate both parity (P) and time
reversal (T) invariance.  If CPT is assumed to be an unbroken
symmetry, a permanent EDM is, then, a signature of CP violation
\cite{bb}.  A non-vanishing permanent EDM has not been measured for
any of the known elementary particles.  In the Standard Model, EDMs
are generated only at the multi-loop level and are predicted to be many
orders of magnitude below the sensitivity of foreseeable
experiments~\cite{hk}.  A non-vanishing EDM therefore would be
unambiguous evidence for CP violation beyond the CKM matrix, and
searches for permanent EDMs of fundamental particles are powerful
probes of extensions of the Standard Model.  In fact, current EDM
bounds are already some of the most stringent constraints on new
physics, and they are highly complementary to many other low energy
constraints, since they require CP violation, but not flavor
violation.

The field of precision muon physics will be transformed in the next
few years~\cite{Hawaiiproc}.  The EDM of the muon is therefore of
special interest. A new BNL experiment~\cite{Semertzidis:1999kv} has
been proposed to measure the muon's EDM at the level of
\begin{equation}
\dmu \sim 10^{-24}~\ecm \ ,
\label{proposedEDM}
\end{equation}
more than five orders of magnitude below the current
bound~\cite{Bailey:1979mn}
\begin{equation}
\dmu = (3.7 \pm 3.4) \times 10^{-19}~\ecm \ ,
\label{currentEDM}
\end{equation}
and even higher precision might be attainable at a future 
neutrino factory complex \cite{Aysto:2001zs}.

The interest in the muon's EDM is further heightened by the recent
measurement of the muon's anomalous magnetic dipole moment (MDM) $\amu
= (g_{\mu}-2)/2$, where $g_{\mu}$ is the muon's gyromagnetic ratio.
The current measurement $\amuexp = 11\ 659\ 202\, (14)\, (6) \times
10^{-10}$~\cite{Brown:2001mg} from the Muon $(g-2)$ Experiment at
Brookhaven differs from the Standard Model prediction
$\amusm$~\cite{Davier:1998si,Marciano:2001qq} by $2.6 \sigma$:
\begin{equation}
\Delta a_\mu \equiv \amuexp - \amusm = (43 \pm 16) \times 10^{-10} \ .
\label{currentamu}
\end{equation}

The muon's EDM and MDM arise from similar operators, and this
tentative evidence for a non-Standard Model contribution to $\amu$
also motivates the search for the muon's EDM~\cite{Feng:2001sq}.  In
fact, the deviation of \eqref{currentamu} may be partially, or even
entirely attributed to a muon EDM! \cite{Feng:2001sq} In
Section~\ref{sec:experimental} we discuss the interplay between the
new physics contributions to the muon MDM and EDM, and their
manifestation in muon dipole moment experiments.  Then in
Section~\ref{sec:theoretical} we present model-independent predictions
for the muon EDM, based on the current $g_\mu-2$ measurement.  Finally
in Section~\ref{sec:susy} we review the theoretical expectations for
the size of the muon EDM in supersymmetry.

\section{Interpretation of Muon Dipole Experiments}
\label{sec:experimental}

Modern measurements of the muon's MDM exploit the equivalence of
cyclotron and spin precession frequencies for $g=2$ fermions
circulating in a perpendicular and uniform magnetic field.
Measurements of the anomalous spin precession frequency are therefore
interpreted as measurements of $\amu$.

The spin precession frequency also receives contributions from the
muon's EDM, however. For a muon traveling with velocity $\bold{\beta}$
perpendicular to both a magnetic field $\bold{B}$ and an electric
field $\bold{E}$, the anomalous spin precession vector is
\begin{eqnarray}
\bold{\omega}_a &=& -a_{\mu} \frac{e}{m_{\mu}} \bold{B}
- d_{\mu} \frac{2c}{\hbar} \bold{\beta} \times \bold{B}
- d_{\mu} \frac{2}{\hbar} \bold{E} \nonumber \\
&&- \frac{e}{m_{\mu}c} \left(\frac{1}{\gamma^2-1} - a_{\mu}\right) 
\bold{\beta} \times \bold{E} \ . 
\label{omega}
\end{eqnarray}
In recent experiments, the last term of \eqref{omega} is removed by
running at the `magic' $\gamma \approx 29.3$, and the third term is
negligible.  For highly relativistic muons with $|\bold{\beta}|
\approx 1$, then, the anomalous precession frequency is found from
\begin{equation}
{|\bold{\omega}_a| \over |\bold{B}|}
\approx \left[ \left( \frac{e}{m_{\mu}} \right)^2
\left(\amusm + \amunp \right)^2 + 
\left(\frac{2c}{\hbar}\right)^2 {\dmunp}^2 \right]^{1/2} \ ,
\label{both}
\end{equation}
where NP denotes new physics contributions, and we have assumed
$\dmunp \gg \dmusm$.

\begin{figure}[t]
\includegraphics[height=2.3in]{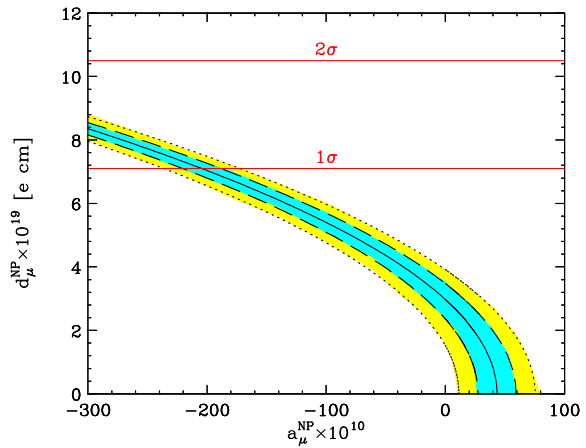}%
\caption{Regions in the $(\amunp,\dmunp)$ plane that are consistent
with the observed $|\bold{\omega}_a|$ at the 1$\sigma$ and 2$\sigma$
levels.  The current 1$\sigma$ and 2$\sigma$ bounds on
$\dmunp$~\protect\cite{Bailey:1979mn} are also shown.}
\label{fig:amu_dmu}
\end{figure}

The observed deviation from the Standard Model prediction for
$|\bold{\omega}_a|$ has been assumed to arise entirely from a MDM and
has been attributed to a new physics contribution of size $\Delta
a_\mu$.  However, from \eqref{both}, we see that, more generally, it
may arise from some combination of magnetic and electric dipole
moments from new physics.  More quantitatively, the effect can also be
due to a combination of new physics MDM and EDM contributions
satisfying
\begin{eqnarray}
\left| \dmunp \right| &\approx& \frac{\hbar e}{2 m_{\mu} c} \,
\sqrt{\ 2\, \amusm\, 
\left(\Delta a_\mu - \amunp  \right)}  \nonumber \\
&\approx& 3.0 \times 10^{-19}~\ecm\ 
\sqrt{1 - \frac{\amunp}{43 \times 10^{-10}}} \ ,
\label{mdmisedm}
\end{eqnarray}
where we have taken into account that $\amunp \ll \amusm$ and
normalized $\amunp$ to the current central value given in
\eqref{currentamu}. In Fig.~\ref{fig:amu_dmu} we show the regions in
the $(\amunp,\dmunp)$ plane that are consistent with the observed
deviation in $|\bold{\omega}_a|$. The current 1$\sigma$ and 2$\sigma$
upper bounds on $\dmunp$~\cite{Bailey:1979mn} are also shown.  We see
that a large fraction of the region allowed by both the current
$g_\mu-2$ measurement \eqref{currentamu} and the $d_\mu$ bound
\eqref{currentEDM} is already within the sensitivity of phase I of the
newly proposed experiment (with sensitivity $\sim 10^{-22}\ \ecm$).

In fact, the observed anomaly may, in principle, be due entirely to
the muon's EDM! This is evident from Eqs.~(\ref{currentEDM}) and
(\ref{mdmisedm}), or from Fig.~\ref{fig:amu_dmu}.  Alternatively, in
the absence of fine-tuned cancellations between $\amunp$ and $\dmunp$,
{\em the results of the Muon $(g-2)$ Experiment also provide the most
stringent bound on $\dmu$ to date}, with $1\sigma$ and $2\sigma$ upper
limits
\begin{eqnarray}
\Delta a_{\mu} &<& 59 \ (75) \times 10^{-10} \Longrightarrow \nonumber \\
\left| \dmunp \right| &<& 3.5 \ (3.9) \times 10^{-19}~\ecm \ .
\label{newdmubound}
\end{eqnarray}

Of course, the effects of $\dmu$ and $\amu$ are physically
distinguishable: while $\amu$ causes precession around the magnetic
field's axis, $\dmu$ leads to oscillation of the muon's spin above and
below the plane of motion.  This oscillation is detectable in the
distribution of positrons from muon decay, and further analysis of the
recent $\amu$ data should tighten the current bounds on $\dmu$
significantly.  Such analysis is currently in progress~\cite{lee} and
should be able to further restrict the allowed region depicted in
Fig.~\ref{fig:amu_dmu}.

The proposed dedicated muon EDM experiment will use a different setup
from the one described above, by applying a constant radial electric
field. As can be seen from \eqref{omega}, the anomalous precession
frequency will then have both a radial component,
\begin{equation}
- d_{\mu} \frac{2c}{\hbar} \bold{\beta} \times \bold{B}
- d_{\mu} \frac{2}{\hbar} \bold{E} \ ,
\label{omega_rad}
\end{equation}
and a vertical component, 
\begin{equation}
-a_{\mu} \frac{e}{m_{\mu}} \bold{B}
- \frac{e}{m_{\mu}c} \left(\frac{1}{\gamma^2-1} - a_{\mu}\right) 
\bold{\beta} \times \bold{E} \ . 
\label{omega_ver}
\end{equation}
Then for any given $\gamma$, and {\em assuming the SM value for
$\amu$}, the electric field can be tuned to cancel the precession from
\eqref{omega_ver} due to $a_\mu$.  The remaining radial component of
$\bold{\omega}_a$ will lead to an oscillating up-down asymmetry in the
counting rate.  Measurements of both the asymmetry and the
spin precession frequency can be used to deduce a limit on $\dmunp$.

As in the $g_{\mu}-2$ experiment, however, the measurement of the spin
precession frequency in the muon EDM experiment receives, in
principle, contributions from both the muon EDM and MDM. In the
presence of a sizable new physics contribution to $\amu$, the
cancellation in \eqref{omega_ver} is not perfect, leaving a
residual radial component
\begin{equation}
-\amunp \frac{e}{m_{\mu}} 
\left( \bold{B} - \frac{1}{c}\ \bold{\beta} \times \bold{E} \right) \ . 
\label{omega_res}
\end{equation}
{}From Eqs.~(\ref{omega_rad}) and (\ref{omega_res}) we then obtain for
the magnitude of the anomalous precession frequency
\begin{eqnarray}
&&|\bold{\omega}_a|^2
= |\bold{B}|^2 \left[ \left( \amunp \frac{e}{m_{\mu}} \right)^2
\left( 1 - \frac{\amusm}{\amusm - \frac{1}{\gamma^2 - 1}} \right)^2
\right. \nonumber \\
&&+ \left. \left( \dmunp \frac{2}{\hbar} \right)^2 
\left( c |\bold{\beta}| + \frac{\amusm}{\frac{|\bold{\beta}|}{c}
\left( \amusm - \frac{1}{\gamma^2 - 1} \right)} \right)^2 \right] ,
\end{eqnarray}
where we have used the tuning condition for \eqref{omega_ver} to
eliminate the electric field.  In the setup of the proposed
experiment, $\gamma \approx 5$, and we can approximate $|\bold{\beta}|
\approx 1 \gg 1/(\gamma^2-1) \gg \amusm$ to get
\begin{equation}
|\bold{\omega}_a|^2
\approx |\bold{B}|^2 \left[ 
   \left( \frac{e}{m_{\mu}} \, \amunp \right)^2
+  \left( \frac{2c}{\hbar}  \, \dmunp  \right)^2 \right]\ .
\end{equation}
We see that the measurement of $\bold{\omega}_a$ again constrains only
a combination (albeit a different one --- cf. \eqref{both}) of
$\amunp$ and $\dmunp$.  This time, the constraint contours are
ellipses centered on the origin in Fig.~\ref{fig:amu_dmu}.  Only by
combining both measurements can the muon EDM and MDM be determined
unambiguously.  Of course, the up-down asymmetry is CP-violating, and
so provides unambiguous information about $\dmunp$ without
contamination from $\amunp$.  The measurement of the up-down asymmetry
is therefore extremely valuable.

\section{Implications of the \bm{$\lowercase{g}_{\mu}-2$} result for 
the Muon's EDM}
\label{sec:theoretical}

The muon's EDM and anomalous MDM are defined through (here and below
we set $\hbar = c = 1$)
\begin{eqnarray}
\label{EDMoperator}
{\cal L}_{\text{EDM}} &=& 
-\frac{i}{2} \dmu \, \bar{\mu} \sigma^{mn} \gamma_5 \mu \, F_{mn} \\ 
{\cal L}_{\text{MDM}} &=& 
\amu \frac{e}{4m_\mu} \, \bar{\mu} \sigma^{mn} \mu \, F_{mn} \ ,
\end{eqnarray}
where $\sigma^{mn} = \frac{i}{2} \left[ \gamma^m, \gamma^n \right]$
and $F$ is the electromagnetic field strength.

\begin{figure}[t]
\includegraphics[height=2.3in]{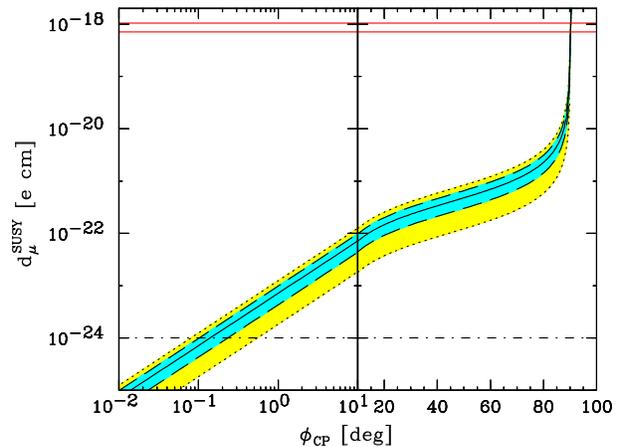}%
\caption{Regions of the $(\phicp, \dmunp)$ plane allowed by the
measured central value of $|\bold{\omega}_a|$ (solid) and its
1$\sigma$ and 2$\sigma$ preferred values (shaded).  The horizontal
dot-dashed line marks the proposed experimental sensitivity to $\dmunp$. 
The red horizontal solid lines denote the current 1$\sigma$ and 2$\sigma$
bounds on $\dmunp$~\protect\cite{Bailey:1979mn}.}
\label{fig:dmu_phi}
\end{figure}

These operators are closely related. Assuming that they
have the same origin, it is useful to write the 
new physics contributions to their coefficients as
\begin{eqnarray}
\dmunp &=& \frac{e}{2m_\mu}\ \Im A \label{imA}\ , \\
\amunp &=& \Re A \ ,                  \label{reA}
\end{eqnarray}
with $A \equiv |A|e^{i\phicp}$.  This defines an experimentally 
measurable quantity $\phicp$ which quantifies the amount of CP
violation in the new physics, independently of its energy scale.
Upon eliminating $|A|$, we find
\begin{equation}
\dmunp = 4.0 \times 10^{-22}~\ecm\ \frac{\amunp}{43 \times 10^{-10}} 
\ \tan\phicp \ .
\label{phicp}
\end{equation}
The measured discrepancy in $|\bold{\omega}_a|$ then constrains
$\phicp$ and $\dmunp$. Eliminating $\amunp$ from Eqs.~(\ref{both}) and
(\ref{phicp}), we find
\begin{eqnarray}
&&\left| \dmunp \right| = \frac{e}{2 m_{\mu}} \,
\amusm \sin\phicp \Biggl[ \ - \ \cos\phicp  \nonumber \\
&&+ \left( \cos^2\phicp + 
{(2\amusm+\Delta a_\mu) \Delta a_\mu \over (\amusm) ^2} \right)^{1/2}  
\Biggr]   \ , 
\end{eqnarray}
The preferred regions of the $(\phicp,\dmunp)$ plane are shown in
Fig.~\ref{fig:dmu_phi}.  For `natural' values of $\phicp \sim 1$,
$\dmunp$ is of order $10^{-22}~\ecm$.  With the proposed $\dmunp$
sensitivity of \eqref{proposedEDM}, all of the 2$\sigma$ allowed
region with $\phicp > 10^{-2} \ {\rm rad}$ yields an observable
signal.

At the same time, while this model-independent analysis indicates that
natural values of $\phicp$ prefer $\dmunp$ well within reach of the
proposed muon EDM experiment, very large values of $\dmunp$ also
require highly fine-tuned $\phicp$.  For example, we see from
Fig.~\ref{fig:dmu_phi} that values of $\dmunp\gsim 10^{-20}\ \ecm$ are
possible only if $|\pi/2 - \phicp| \sim 10^{-3}$.  This is a
consequence of the fact that EDMs are CP-odd and $\dmusm \approx 0$,
and so $\dmunp$ appears only quadratically in $|\bold{\omega}_a|$.
Without a strong motivation for $\phicp \approx \pi/2$, it is
therefore natural to expect the EDM contribution to
$|\bold{\omega}_a|$ to be negligible.

\section{Theoretical Expectations for \bm{$\lowercase{d}_\mu$} in 
Supersymmetry}
\label{sec:susy}

Our discussion up to now has been completely model-independent.  In
specific models, however, it may be difficult to achieve values of
$\dmu$ large enough to saturate the bound of \eqref{newdmubound}.  For
example, in supersymmetry, assuming flavor conservation and taking
extreme values of superparticle masses ($\sim 100~\gev$) and $\tb$
($\tb \sim 50$) to maximize the effect, the largest possible value of
$\amu$ is $a_{\mu}^{\text{max}} \sim 10^{-7}$~\cite{Feng:2001tr}.
Very roughly, one therefore expects a maximal $d_{\mu}$ of order $(e
\hbar / 2 m_{\mu} c) a_{\mu}^{\text{max}} \sim 10^{-20}~\ecm$ in
supersymmetry.

With additional model assumptions, however, it is possible to further
narrow down the expected range of $\dmunp$ in supersymmetry.  The EDM
operator of \eqref{EDMoperator} couples left- and right-handed muons,
and so requires a mass insertion to flip the chirality.  The natural
choice for this mass is the lepton mass.  On dimensional grounds, one
therefore expects
\begin{equation}
\label{massscaling} 
\dmunp \propto \frac{m_{\mu}}{\tilde{m}^2}\ , 
\end{equation} 
where $\tilde{m}$ is the mass scale of the new physics. If the new
physics is flavor blind, $d_f \propto m_f$ for all fermions $f$, which
we refer to as `naive scaling.'  In particular,
\begin{equation}\label{naive}
d_\mu \approx {m_\mu\over m_e}\, d_e \ .
\end{equation}

The current bound on the electron EDM is $d_e = 1.8\, (1.2)\, (1.0)
\times 10^{-27}~\ecm$~\cite{Commins:1994gv}.  Combining the
statistical and systematic errors in quadrature, this bound and
\eqref{naive} imply
\begin{equation}
d_\mu \alt 9.1\times 10^{-25}~\ecm \ ,
\label{muedmlimit}
\end{equation}
at the 90\% CL, barely below the sensitivity of \eqref{proposedEDM}.
Naive scaling must be violated if a non-vanishing $\dmu$ is to be
observable at the proposed experiment.  On the other hand, the
proximity of the limit of \eqref{muedmlimit} to the projected
experimental sensitivity of \eqref{proposedEDM} implies that even
relatively small departures from naive scaling may yield an observable
signal.

Is naive scaling violation well-motivated, and can the violation be
large enough to produce an observable EDM for the muon? To investigate
these questions quantitatively, we consider supersymmetry
\cite{recentwork}.  (For violations of naive scaling in other models,
see, for example, Ref.~\cite{Babu:2000cz}.)  Many additional mass
parameters are introduced in supersymmetric extensions of the Standard
Model. These are in general complex and are new sources of CP
violation, leading to a separate, major challenge for SUSY model
building along with flavor violation. For a recent discussion of the
supersymmetric CP problem in various supersymmetry breaking schemes,
see Ref.~\cite{Dine:2001ne}.

In the minimal supersymmetric model, naive scaling requires

\noindent $\bullet$ Degeneracy: Generation-independent slepton masses.

\noindent $\bullet$ Proportionality: The $A$ terms must scale with the 
corresponding fermion mass.

\noindent $\bullet$ Flavor conservation: Vanishing off-diagonal elements
for the sfermion masses and the $A$-terms.

We now briefly discuss violations of each of these properties in turn.

Scalar degeneracy is the most obvious way to reduce flavor changing
effects to allowable levels. Therefore many schemes for mediating
supersymmetry breaking try to achieve degeneracy.  However, in many of
these, with the exception of simple gauge mediation models, there may
be non-negligible contributions to scalar masses that are
generation-dependent. For example, scalar non-degeneracy is typical in
alignment models~\cite{Nir:1993mx} or models with anomalous U(1)
contributions to the sfermion masses where the sfermion hierarchy is
often inverted relative to the fermion mass hierarchy~\cite{models}.

We now consider a simple model-independent parameterization to explore
the impact of non-degenerate selectron and smuon masses.  We set
$m_{\tilde{e}_R} = m_{\tilde{e}_L} = \msel$ and $m_{\tilde{\mu}_R} =
m_{\tilde{\mu}_L} = \msmu$ and assume vanishing $A$ parameters.  For
fixed values of $M_1$, $M_2$, $|\mu|$, and large $\tb$, then, to a
good approximation both $\de$ and $\dmu$ are proportional to $\sin
\phicp\, \tb$, and we assume that $\sin \phicp\, \tb$ saturates the
$\de$ bound.

\begin{figure}[tbp]
\includegraphics[height=2.3in]{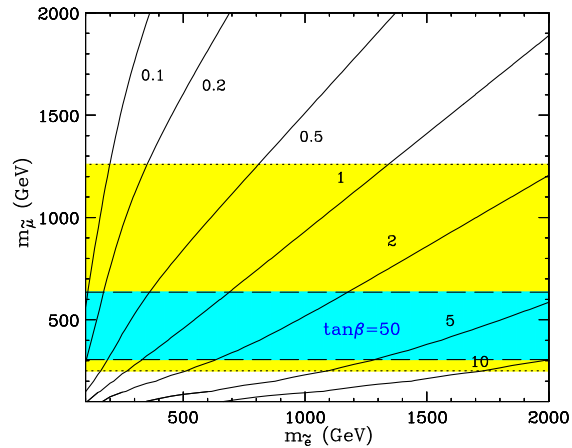}%
\caption{Contours of $\dmu \times 10^{24}$ in $\ecm$ for varying
$m_{\tilde{e}_R} = m_{\tilde{e}_L} = \msel$ and $m_{\tilde{\mu}_R} =
m_{\tilde{\mu}_L} = \msmu$ for vanishing $A$ terms, fixed $|\mu| =
500~\gev$ and $M_2 = 300~\gev$, and $M_1 = (g_1^2/g_2^2) M_2$
determined from gaugino mass unification.  The CP-violating phase
is assumed to saturate the bound $\de < 4.4 \times 10^{-27}~\ecm$. 
The shaded regions are preferred by $\amu$ at
$1\sigma$ and $2\sigma$ for $\tb=50$. }
\label{fig:msel_msmu}
\end{figure}

Contours of $\dmu$ are given in Fig.~\ref{fig:msel_msmu}.  Observable
values of $\dmu$ are possible even for small violations of
non-degeneracy; for example, for $\msmu/\msel \alt 0.9$, muon EDMs
greater than $10^{-24}~\ecm$ are possible. The current value of $\amu$
also favors light smuons and large EDMs.  The smuon mass regions
preferred by the current $\amu$ anomaly are given in
Fig.~\ref{fig:msel_msmu} for $\tb = 50$.  Within the $1\sigma$
preferred region, $\dmu$ may be as large as $4\ (10) \times
10^{-24}~\ecm$ for $\msel < 1\ (2)~\tev$.  Our assumed value of $\tb$
is conservative; for smaller $\tb$, the preferred smuon masses are
lower and the possible $\dmu$ values larger.

Naive scaling is also broken if the $A$-terms are not proportional to
the corresponding Yukawa couplings.  Just as in the case of
non-degeneracy, deviations from proportionality are found in many
models. Although for large $\tb$, the $A$ term contribution to the EDM
is suppressed relative to the typically dominant chargino
contribution,
there are many possibilities that may yield large effects.  In
Ref.~\cite{Ibrahim:2001jz}, for example, it was noted that $A_e$ may
be such that the chargino and neutralino contributions to $d_e$
cancel, while, since $A_e\neq A_\mu$, there is no cancellation in
$d_\mu$, and observable values are possible.

Finally, most models of high-scale supersymmetry
breaking~\cite{Dine:2001ne} typically contain flavor violation as
well. In particular, smuon-stau mixing leads to a potentially
significant enhancement in $\dmu$, because it breaks naive scaling by
introducing contributions enhanced by ${m_\tau\over m_\mu}$. In order
to evaluate the significance of this enhancement, we must first
determine how large the flavor violation may be.  Taking into account
the current $\tau \to \mu \gamma$ constraint, we found that values of
$\dmunp$ as large as $10^{-22}\ecm$ are possible \cite{Feng:2001sq}.

In conclusion, the proposal to measure the muon EDM at the level of
$10^{-24}~\ecm$ potentially improves existing sensitivities by five
orders of magnitude.  While the existing deviation in $g_{\mu}-2$ may
be interpreted as evidence for new physics in either the muon's MDM or
EDM, the proposed experiment will definitively resolve this ambiguity,
and may also uncover new physics in a wide variety of superysmmetric
extensions of the Standard Model.

\vspace{-0.4cm}
\begin{acknowledgments}
The work of J.L.F. is supported in part by the U.~S.~Department of
Energy under cooperative research agreement DF--FC02--94ER40818.
K.T.M. thanks the Fermilab Theory Group, 
the ANL Theory Group 
and the Aspen Center for Physics for hospitality during 
the completion of this work.
\end{acknowledgments}


\end{document}